\documentclass{IEEEtran}
\usepackage{cite}
\usepackage{amsmath,amssymb,amsfonts}
\usepackage{algorithmic}
\usepackage{graphicx}
\usepackage{textcomp}
\usepackage{soul}
\def\BibTeX{{\rm B\kern-.05em{\sc i\kern-.025em b}\kern-.08em
    T\kern-.1667em\lower.7ex\hbox{E}\kern-.125emX}}
\begin{document}

\onecolumn % switch to one column
\pagestyle{empty} % don't print page number
\begin{center}
  \large\bfseries  % font for your notice
  This work has been submitted to the IEEE for possible publication. Copyright may be transferred without notice, after which this version may no longer be accessible.
\end{center}
\twocolumn % switch to 2 columns
\setcounter{page}{1} % page number for start of manuscript
% \pagestyle[headings} % per your code

% \title{IMPULSE: Digital Compute-in-Memory Macro for Low-Power Spiking Neural Inference}
\title{IMPULSE: A 65nm Digital Compute-in-Memory Macro with Fused Weights and Membrane Potential for Spike-based Sequential Learning Tasks}

\author{Amogh Agrawal$^*$, Mustafa Ali$^*$, Minsuk Koo, Nitin Rathi, Akhilesh Jaiswal, Kaushik Roy, \IEEEmembership{Fellow, IEEE}, 
\thanks{All authors are with School of Electrical and Computer Engineering, Purdue University, West Lafayette, IN 47906 USA. (e-mail: agrawa64@purdue.edu)}
\thanks{*Equal contribution}
}

\maketitle

\begin{abstract}

The inherent dynamics of the neuron membrane potential in Spiking Neural Networks (SNNs) allows processing of sequential learning tasks, avoiding the complexity of recurrent neural networks. The highly-sparse spike-based computations in such spatio-temporal data can be leveraged for energy-efficiency. However, the membrane potential incurs additional memory access bottlenecks in current SNN hardware. To that effect, we propose a 10T-SRAM compute-in-memory (CIM) macro, specifically designed for state-of-the-art SNN inference. It consists of a fused weight (W\textsubscript{MEM}) and membrane potential (V\textsubscript{MEM}) memory and inherently exploits sparsity in input spikes leading to $\sim$97.4\% reduction in energy-delay-product (EDP) at 85\% sparsity (typical of SNNs considered in this work) compared to the case of no sparsity. We propose staggered data mapping and reconfigurable peripherals for handling different bit-precision requirements of W\textsubscript{MEM} and V\textsubscript{MEM}, while supporting multiple neuron functionalities. The proposed macro was fabricated in 65nm CMOS technology, achieving an energy-efficiency of 0.99TOPS/W at 0.85V supply and 200MHz frequency for signed 11-bit operations. We evaluate the SNN for sentiment classification from the IMDB dataset of movie reviews and achieve within $\sim$1\% accuracy of an LSTM network with $8.5\times$ lower parameters.
%% Nitin

\end{abstract}


\vspace{-1mm}
\section{Introduction}

\begin{figure}[t]
    \centering
    \includegraphics[width=\linewidth]{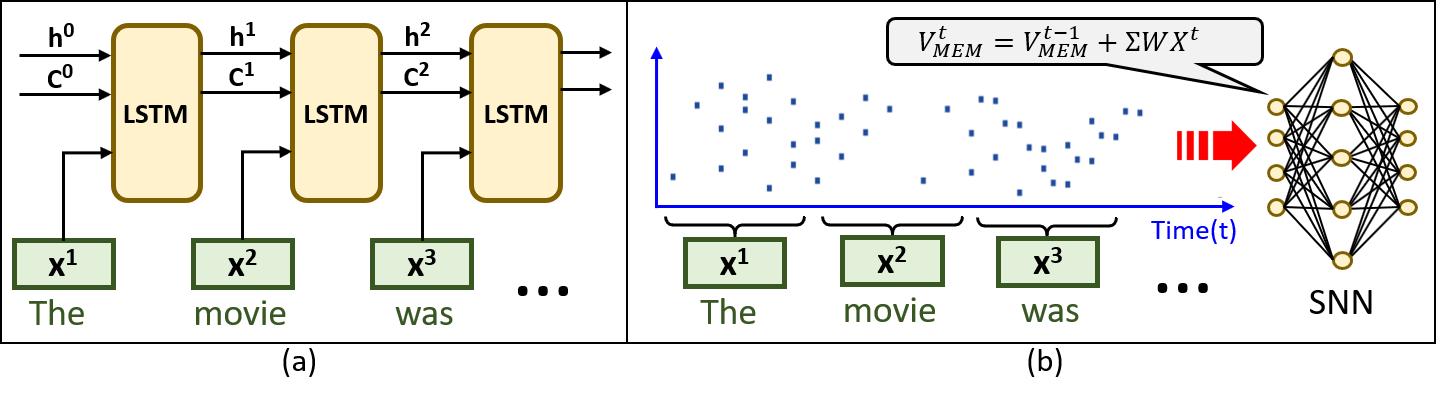}
    \vspace{-10mm}
    \caption{(a) LSTM having hidden state (h\textsuperscript{t}, C\textsuperscript{t}) for processing sequential tasks. (b) Intrinsic temporal dynamics of neuron membrane potentials (V\textsuperscript{t}\textsubscript{MEM}) in SNNs for processing sequential tasks.}
    \vspace{-4mm}

    \label{fig:snn_dynamical}
\end{figure}

Spiking Neural Networks (SNNs) aim to harness the inherent energy-efficiency arising from highly sparse and event-driven spike-based information processing \cite{davies2018loihi,merolla2014million}. SNN algorithms continue to develop rapidly, achieving image classification accuracies close to state-of-the-art \cite{rathi2020diet}. The SNN approach uses binary inputs and outputs (1$-$spike, 0$-$no spike) over several timesteps, unlike traditional ANNs. More importantly, the neuron membrane potential plays a key role which defines each neuron's current state, thereby allowing SNNs to process  dynamical (temporal) aspects in the data. This makes them suited for sequential learning tasks, avoiding the complexity of recurrent neural networks, such as Long Short-Term Memory (LSTM). Fig.~\ref{fig:snn_dynamical} illustrates the processing of sequential inputs in both LSTM and SNN. In LSTM, the hidden state ($h,C$) stores information about all past inputs the network has seen before \cite{hochreiter1997long}. To enable this, the previous hidden state is fed back as input through a recurrent connection along with the current input. Whereas, in SNNs the inherent recurrence in membrane potential acts as a memory to store information about past inputs. Each LSTM layer has $4mn+n^2$ parameters, compared to $mn$ parameters in an SNN layer, where $m$ and $n$ are input and output dimensions, respectively. 
%% Nitin

\begin{figure}[t]
    \centering
    \includegraphics[width=\linewidth]{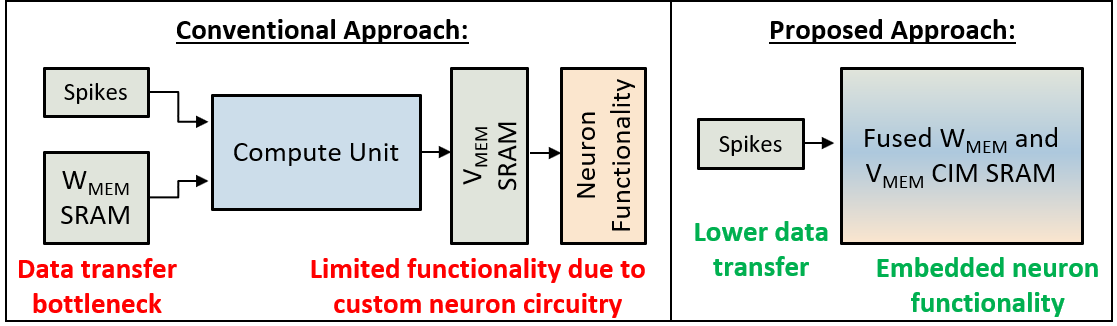}
    \vspace{-8mm}
    \caption{Limitations of current digital SNN hardware accelerators and our proposed approach of fused weight and membrane potential CIM SRAM.}
    \vspace{-4mm}    
    \label{fig:motiv}
\end{figure}

\begin{figure*}[t]
    \centering
    \includegraphics[width=\linewidth]{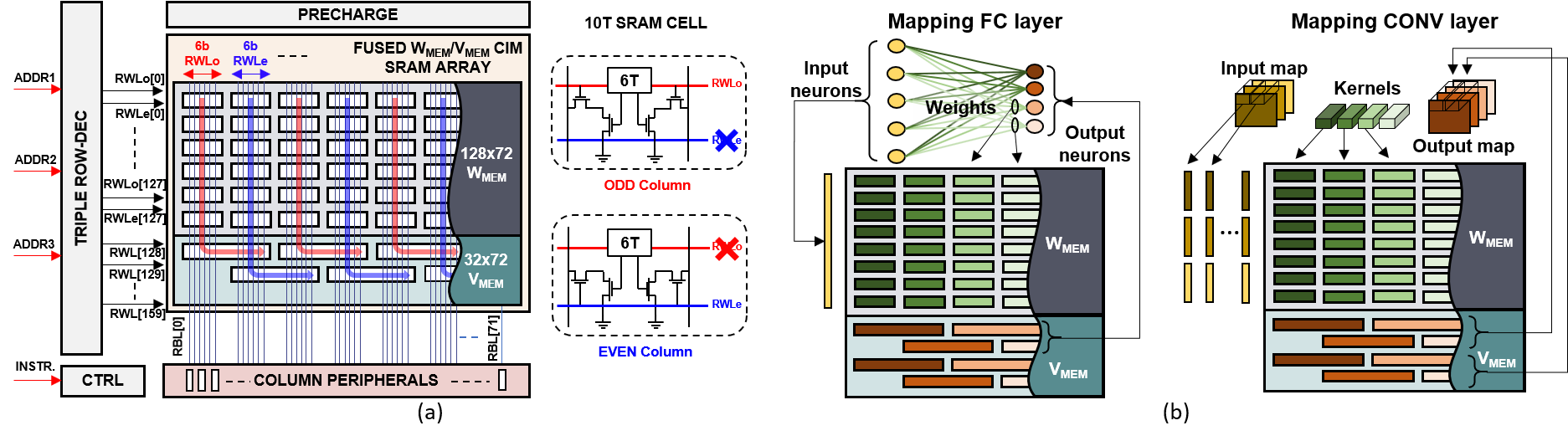}
    \vspace{-10mm}    
    \caption{(a) Organization of the fused W\textsubscript{MEM}/V\textsubscript{MEM} 10T-SRAM macro. (b) Mapping of FC and Conv layers on the proposed macro.}
    \vspace{-6mm}    
    \label{fig:arch}
\end{figure*}

However, processing of membrane potential over several timesteps incurs additional memory-access bottlenecks, specific to SNNs \cite{narayanan2020spinalflow}. The main challenges in current SNN hardware accelerators are: (1) Significant energy consumption due to data movements from weight and membrane potential SRAMs to compute units. (2) They support limited SNN functionality due to the adoption of custom neuron circuitry, which are power and area expensive, making them restricted to simpler tasks. These are illustrated in Fig.~\ref{fig:motiv}.

To overcome these challenges, we propose: (1) an SRAM-based CIM macro which integrates all instructions required for SNN inference such as accumulate, thresholding, spike-check and reset, within the fused W\textsubscript{MEM} and V\textsubscript{MEM} memory, thereby reducing on-chip SRAM accesses. (2) Support for multiple types of neurons through the same in-memory instructions $–$ integrate-fire (IF), leaky-IF (LIF), and residual membrane potential (RMP) \cite{han2020rmp} neurons. (3) A staggered data mapping and reconfigurable column peripherals for maintaining different bit-precision requirements of W\textsubscript{MEM} and V\textsubscript{MEM}, while allowing full utilization of the array and column peripherals.

%The proposed macro was fabricated in 65nm CMOS, consisting of a fused 9Kb W\textsubscript{MEM} and 2.25Kb V\textsubscript{MEM} (11.25Kb total) 10T SRAM CIM array. The chip 

\vspace{-1mm}
\section{IMPULSE: Structure and Operation}

Fig.~\ref{fig:arch}(a) shows the overall organization of the macro. Each of the 128 rows in the W\textsubscript{MEM} subarray corresponds to an input neuron, storing twelve 6-bit signed weights. Each row has two read wordlines (RWLo/RWLe) and the weights are interleaved, such that the first six bits are on RWLo, next six on RWLe, and so on. In each cycle, either of RWLo or RWLe are enabled. The V\textsubscript{MEM} subarray contains 32 rows with single RWL, each row storing six signed values. The V\textsubscript{MEM} corresponding to odd and even weight columns are stored in different rows with a staggered alignment. This mapping technique compactly handles the different precision requirements for weights (6-bit) and V\textsubscript{MEM} (11-bit), with full utilization of all column peripherals in each odd/even cycle. The two subarrays are fused through common bitlines. The triple-row decoder can take three addresses and enables two RWLs and one WWL simultaneously. Fig.~\ref{fig:arch}(b) illustrates how fully-connected (FC) and convolutional (Conv) layers can be mapped to the macro.

\vspace{-3mm}
\subsection{Reconfigurable Column Peripherals}

\begin{figure}[t]
    \centering
    \includegraphics[width=\linewidth]{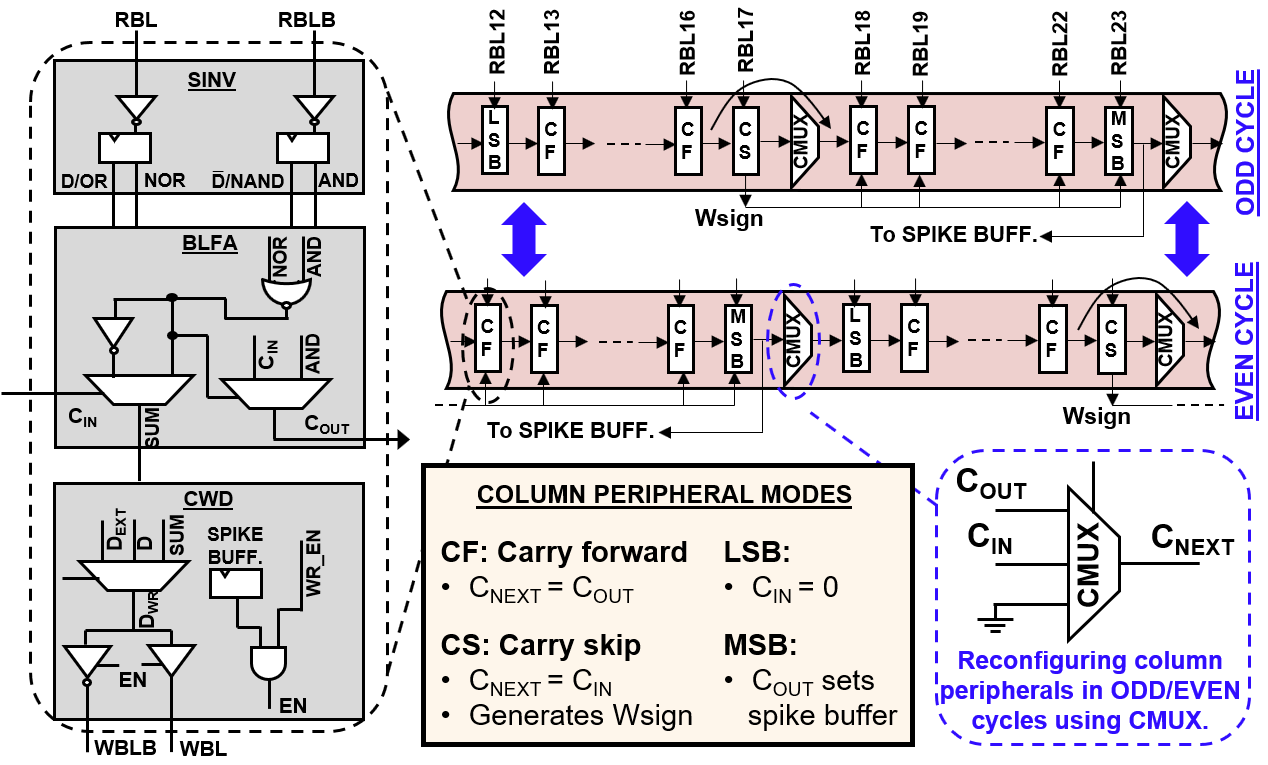}
    \vspace{-8mm}
    \caption{Detailed description of the reconfigurable column peripherals.}
    \vspace{-4mm}
    \label{fig:peri}
\end{figure}

Fig.~\ref{fig:peri} describes the column peripherals in detail. Each set of bitlines (RBL,RBLB,WBL,WBLB) connects to a column peripheral circuitry. During the CIM operations, the RBL gives NOR/OR, while RBLB gives NAND/AND, of the data from the two enabled RWL rows. This is sensed and latched using the sensing inverters (SINV). The bitwise logic full adder (BLFA) is designed to generate SUM and COUT using these bitwise signals. SUM is fed to the conditional write-driver (CWD), to be written into the enabled WWL destination row, while COUT is forwarded to the adjacent column peripheral for the accumulate operation, forming a ripple-carry adder. 

The modular design of the adder using BLFAs from each column peripheral allows reconfigurability during odd/even cycles. To account for the staggered data mapping, each column peripheral can be reconfigured in $–$ carry forward (CF), carry skip (CS), LSB and MSB modes, with the help of Carry-MUXes (CMUX), as shown in Fig.~\ref{fig:peri}. For example, during odd cycle, Col[0-11] form one adder, Col[12-23] forms another, and so on. Whereas during even cycle, Col[6-17] form one adder, Col[18-29] form another, and so on. It's worthwhile to note that the sixth bit of V\textsubscript{MEM} aligns with MSB of the weight (Wsign), and needs to be kept `0' to correctly read Wsign (hence, 11-bit V\textsubscript{MEM}). CS block forwards this Wsign to the next six column peripherals for performing the full 11-bit accumulate operation.

\vspace{-2.5mm}
\subsection{In-Memory SNN Instructions}

\begin{figure}[t]
    \centering
    \includegraphics[width=\linewidth]{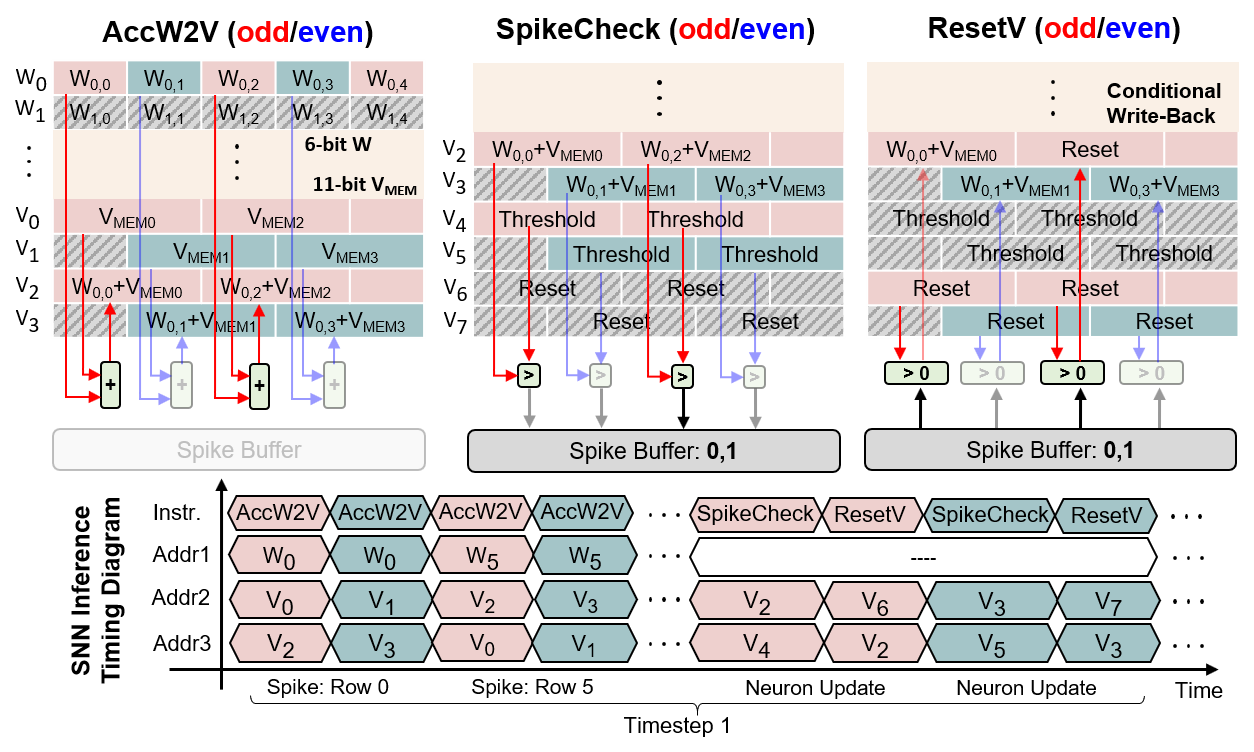}
    \vspace{-9mm}
    \caption{Illustration of supported in-memory SNN instructions.}
    \vspace{-4mm}
    \label{fig:instr}
\end{figure}

Fig.~\ref{fig:instr} shows the supported in-memory SNN instructions. In Accumulate-W-to-V (AccW2V) instruction, depending on which input neuron spikes, the corresponding RWL from the W\textsubscript{MEM} block (RWLo in odd cycle, RWLe in even cycle), and one RWL and one WWL from the V\textsubscript{MEM} block are enabled simultaneously. Thus, 6-bit weights and 11-bit membrane potentials are accumulated and updated simultaneously, along the whole row. Similarly, AccV2V (not shown) accumulates two V\textsubscript{MEM} rows. During the SpikeCheck instruction, two RWLs from the V\textsubscript{MEM} blocks are enabled, one corresponding to the membrane potential to be checked, and other storing the threshold. The adders formed by the column peripherals act as comparators in this case, by checking if the sum is greater or less than 0. This can be achieved by checking the COUT from MSB column peripheral, which is then utilized in setting the corresponding spike buffer if the membrane potential exceeds the threshold. Subsequently, the ResetV instruction follows the SpikeCheck instruction. During ResetV, one RWL and one WWL are enabled in the V\textsubscript{MEM} block corresponding to the reset value and the destination membrane potential, respectively. The BLFA is bypassed in this instruction, and the reset value read in the SINV block gets directly transferred to the CWD block. The spike buffers determine whether the CWD drives the WBLs/WBLBs or leaves them precharged, thereby conditionally writing into only those membrane potentials in a row which have spiked. Thus, during the SNN inference, each input spike translates to AccW2V (odd and even) instruction for accumulating weights to membrane potentials. At the end of each timestep, the neuron output is computed using SpikeCheck and ResetV instructions, thereby generating output spikes. This process is repeated for all timesteps.

\vspace{-2.5mm}
\subsection{Multiple Neuron Functionalities}

\begin{figure}[t]
    \centering
    \includegraphics[width=\linewidth]{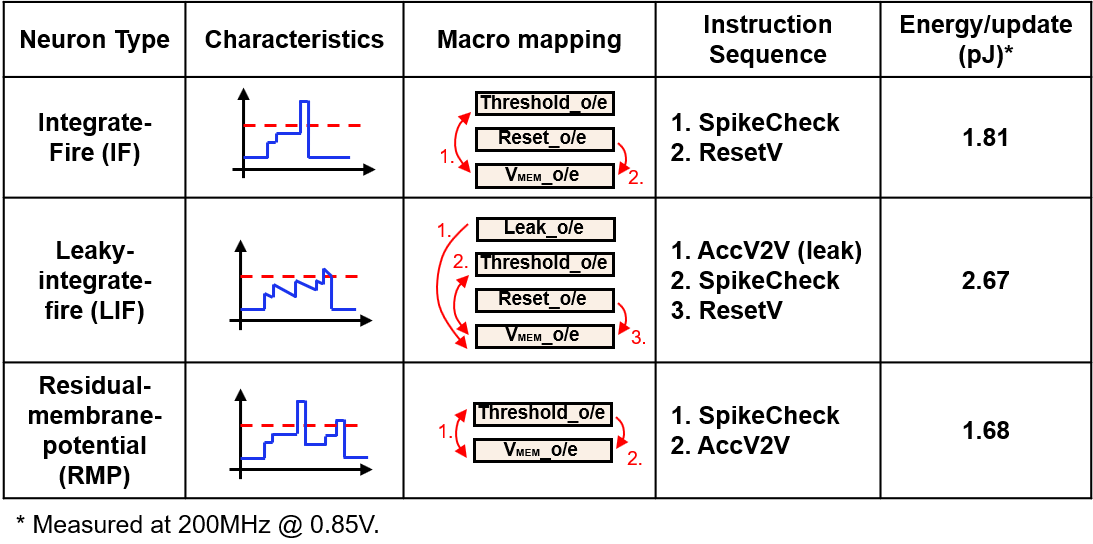}
    \vspace{-8mm}
    \caption{Multiple neurons can be implemented using in-memory instructions.}
    \vspace{-4mm}
    \label{fig:neurons}
\end{figure}

\begin{figure}[t]
    \centering
    \includegraphics[width=\linewidth]{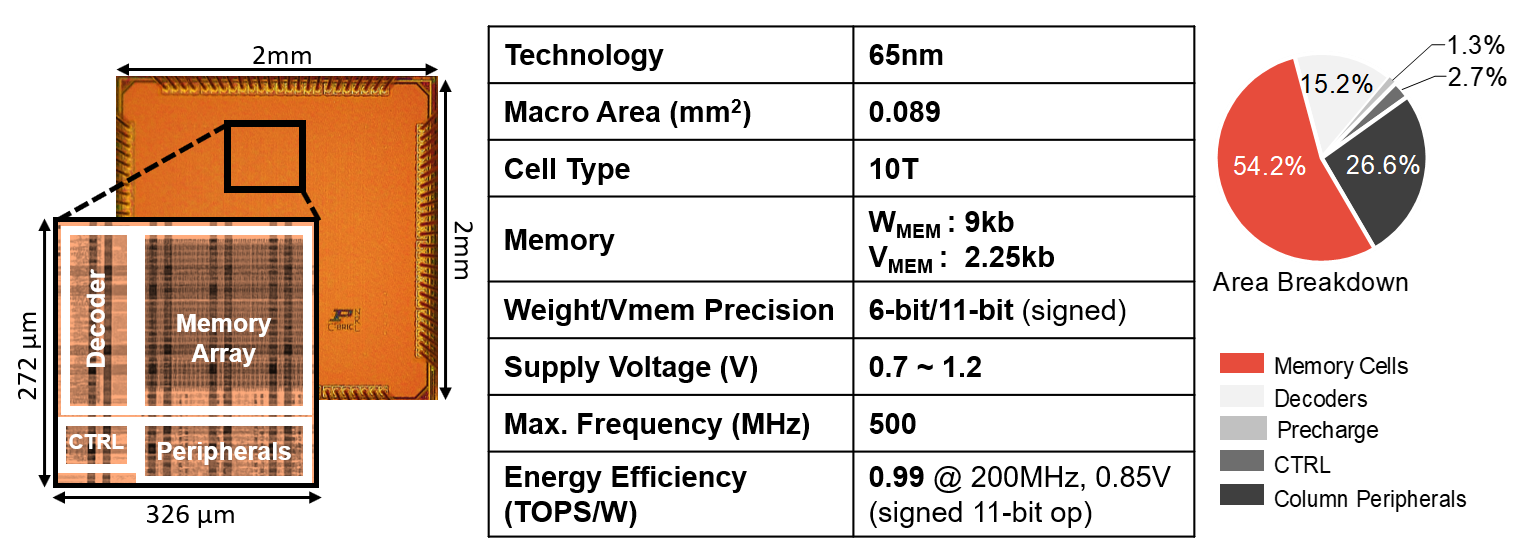}
    \vspace{-8mm}
    \caption{Die micrograph, prototype chip features, and area breakdown.}
    \vspace{-4mm}
    \label{fig:chip_summary}
\end{figure}

While the IF neuron model is predominant in most SNN applications, other neuron functionalities such as LIF and RMP \cite{han2020rmp} have demonstrated superior performance for certain applications \cite{rathi2020diet}. Fig.~\ref{fig:neurons} illustrates the characteristics of these neuron models, and how they can be implemented on IMPULSE through a combination of SpikeCheck, AccV2V, and ResetV instructions. The IF neuron can be implemented simply by using SpikeCheck and ResetV instructions as described in the previous section. The LIF neuron characteristics adds a `leak factor', which can be incorporated by using AccV2V instruction to subtract the `leak' value from the membrane potential, before using SpikeCheck and ResetV instructions. The RMP neuron, on the other hand, uses a soft reset, where the threshold value is subtracted from the membrane potential if it spikes, instead to resetting it. Thus, it can be implemented by using the AccV2V instruction after SpikeCheck, instead of ResetV. The figure also tabulates the measured energy per neuron-update at 200MHz clock and 0.85V supply.

% \section{Sentiment Analysis using SNNs}

\vspace{-1mm}
\section{Implementation Results}

\begin{figure}[t]
    \centering
    \includegraphics[width=\linewidth]{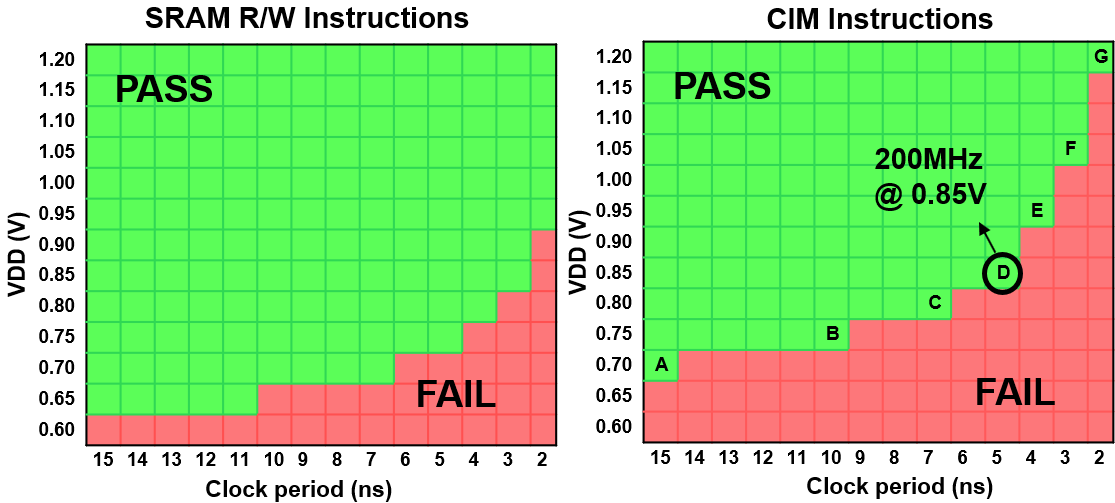}
    \vspace{-8mm}
    \caption{Shmoo plot for read/write and CIM instructions.}
    \vspace{-4mm}
    \label{fig:shmoo}
\end{figure}

\begin{figure}[t]
    \centering
    \includegraphics[width=\linewidth]{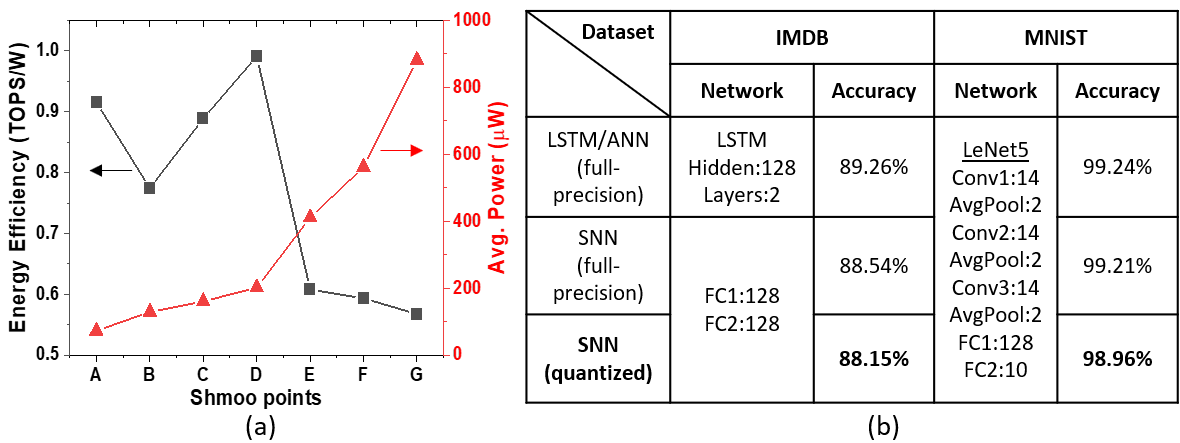}
    \vspace{-8mm}
    \caption{(a) Measured average power and energy-efficiency for AccW2V instruction. (b) SNN architecture and accuracy on IMDB and MNIST datasets.}
    \vspace{-4mm}
    \label{fig:energy_efficiency_accuracy_table}
\end{figure}

The prototype test chip was fabricated in 65nm CMOS technology (Fig.~\ref{fig:chip_summary}).  It achieves a 54.2\% memory area efficiency. Fig.~\ref{fig:shmoo} shows the Shmoo plot for both read/write and CIM instructions. Since CIM instructions are more complex, their operating window is smaller. Note that the CIM Shmoo includes the test of all CIM instructions. Fig.~\ref{fig:energy_efficiency_accuracy_table}(a) plots the measured average power consumption and the energy-efficiency for AccW2V instruction (which is the main synaptic operation), for various possible operating points of interest identified on the CIM Shmoo (A-G). It can be observed that point D (200MHz clock and 0.85V supply) achieves optimal energy-efficiency with 0.99 TOPS/W for AccW2V (1 op = 11-bit operation). Other instructions $-$ AccV2V, ResetV, and SpikeCheck achieve 1.18, 1.02, and 1.22 TOPS/W, respectively, at point D.
%The measured energy/op (1 op = 11-bit operation) at 200MHz and 0.85V supply achieves the optimal energy-efficiency with 1.01, 0.85, 0.98, and 0.82 pJ/op for AccW2V, AccV2V, ResetV, and SpikeCheck, respectively. 
%The area and power breakdown (simulated) of the macro are shown in Fig. \ref{fig:sparsity}(b).

\begin{figure}[t]
    \centering
    \includegraphics[width=\linewidth]{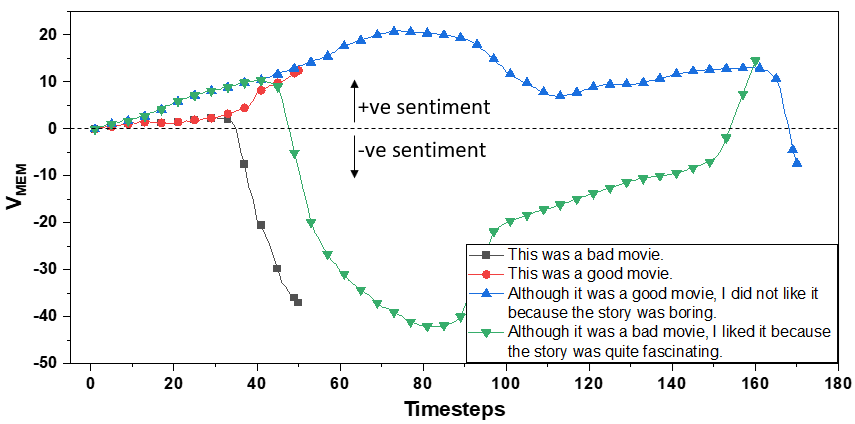}
    \vspace{-8mm}
    \caption{Progression of the final output neuron's membrane potential with timesteps, where each word is presented to the SNN for 10 timesteps.}
    \vspace{-4.5mm}
    \label{fig:vmem}
\end{figure}

We train an SNN with an input layer (100 neurons), two FC layers (128 neurons), and an output layer (1 neuron) to classify movie reviews from IMDB dataset~\cite{maas2011learning}. SNN has 6-bit signed weights, and 11-bit V\textsubscript{MEM} with RMP neurons. Each word in the sentence is converted to a 100-d vector using the Glove model~\cite{pennington2014glove}, and presented to the SNN for 10 timesteps. SNN is trained with surrogate-gradient based backpropagation with threshold and leak optimization~\cite{rathi2020diet}. The input layer acts as spike-encoder and the two FC layers are mapped successively on IMPULSE. The SNN achieves an accuracy of $88.15\%$ with $29.3K$ trainable parameters compared to a 2-layer LSTM network with $247.8K$ parameters (Fig.~\ref{fig:energy_efficiency_accuracy_table}(b)). Fig.~\ref{fig:vmem} shows the dynamics of output layer neuron's membrane potential for exemplary inputs, while processing each word sequentially. V\textsubscript{MEM} value above zero represents positive sentiment and vice-versa.
%% Nitin - .. Shows good accuracy compared to LSTM and 1/8th lower complexity... Vmem plot....
We also trained an SNN with modified LeNet5 architecture to demonstrate image classification from MNIST dataset and achieved 98.96\% accuracy with 10 timesteps. The first Conv layer acts as a spike-encoder, while Conv2,3 and FC1,2 are successively mapped on IMPULSE. Note that input channels for Conv layers were kept 14 with 3$\times$3 kernel size to restrict the fan-in to 128 (3$\times$3$\times$14$=$126), to fit our macro. Similarly, the number of neurons in FC layers was kept $\leq$128.

\begin{figure}[t]
    \centering
    \includegraphics[width=\linewidth]{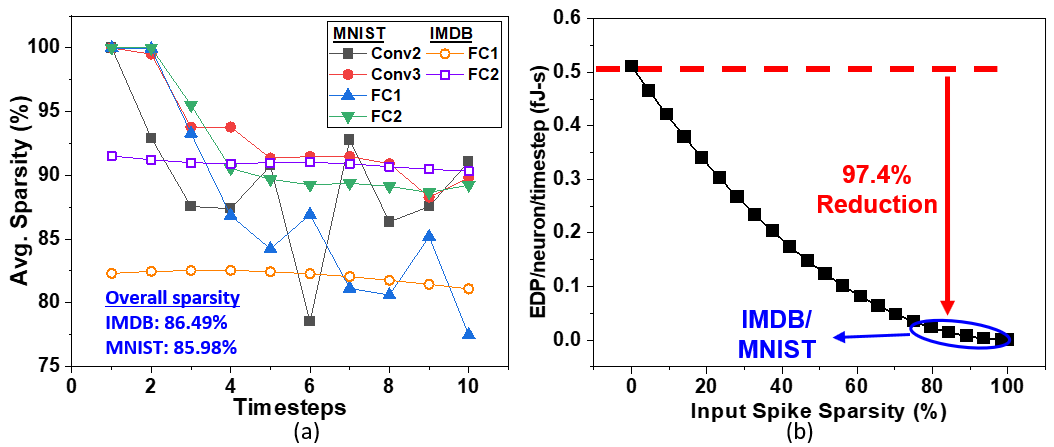}
    \vspace{-8mm}
    \caption{(a) Average sparsity obtained at each layer of SNN for each timestep. (b) Measured EDP per-neuron per-timestep with varying sparsity.}
    \vspace{-4.5mm}
    \label{fig:sparsity}
\end{figure}

Fig.~\ref{fig:sparsity}(a) plots the average sparsity in the spikes observed at each layer for each of the 10 timesteps. Note that these are averaged over each word and each image of the IMDB and MNIST test dataset, respectively. The overall sparsity of $\sim$85\% is achieved in both cases, which leads to significant energy improvements. The proposed macro exploits the input-spike sparsity in SNNs since the number of spikes determine the number and sequence of instructions executed. The measured EDP per-neuron per-timestep is plotted with varying sparsity (0\%: all 128 input neurons spike, 100\%: no input neuron spikes) showing a 97.4\% reduction in EDP at 85\% sparsity, as shown in Fig.~\ref{fig:sparsity}(b). 

%An FC layer (100 input neurons and 10 output neurons) quantized SNN with IF neurons was trained on a scaled MNIST dataset (10x10 pixels) and mapped to the macro as proof-of-concept, achieving 91.33\% classification accuracy for 20 timesteps. The image pixels are converted to spikes through a Poisson generator with firing rate proportional to pixel intensity \cite{rathi2020enabling}. Fig.8 illustrates the inference process and the output spike-count generated by the prototype chip. This was repeated for 10,000 test images and the obtained confusion matrix for the average spike-count is also plotted. 

We compare our proposed approach with other state-of-the-art SNN macros \cite{liu2017scalable,koo2019area,9336139} and digital CIM macros \cite{jeloka2015configurable,wang201914,kim2020z} (Table \ref{fig:comp_table}). Among the SNN macros, \cite{liu2017scalable} has a poor energy efficiency due to time-based digital oscillator circuits for implementing neuron functionality, while \cite{9336139} has 2.7$\times$ lower energy-efficiency (assuming linear scaling with bit-precision) compared to our design due to very low-frequency operation. Ref. \cite{koo2019area} implemented an area-efficient weight memory, however, this design would still suffer from membrane potential bottlenecks. On the other hand, among digital CIM macros, 6T-SRAM based macro \cite{jeloka2015configurable} achieves high memory area efficiency, however, it suffers from read disturb failures. Ref. \cite{wang201914} and \cite{kim2020z} proposed 8T-SRAM based CIM macros developed for ANNs, having 1.5$\times$  and 2.2$\times$ lower energy-efficiency compared to our design, respectively. Thus, our work is the only digital CIM based SNN macro, with support for all instructions required for SNN inference, and multiple neuron functionalities. It also supports both FC and Conv layers and is scalable to larger networks by employing a distributed multi-macro architecture.

\begin{table}[t]
    \centering
    \caption{Comparison with other SNN and CIM macros.}
    \vspace{-2mm}
    \includegraphics[width=\linewidth]{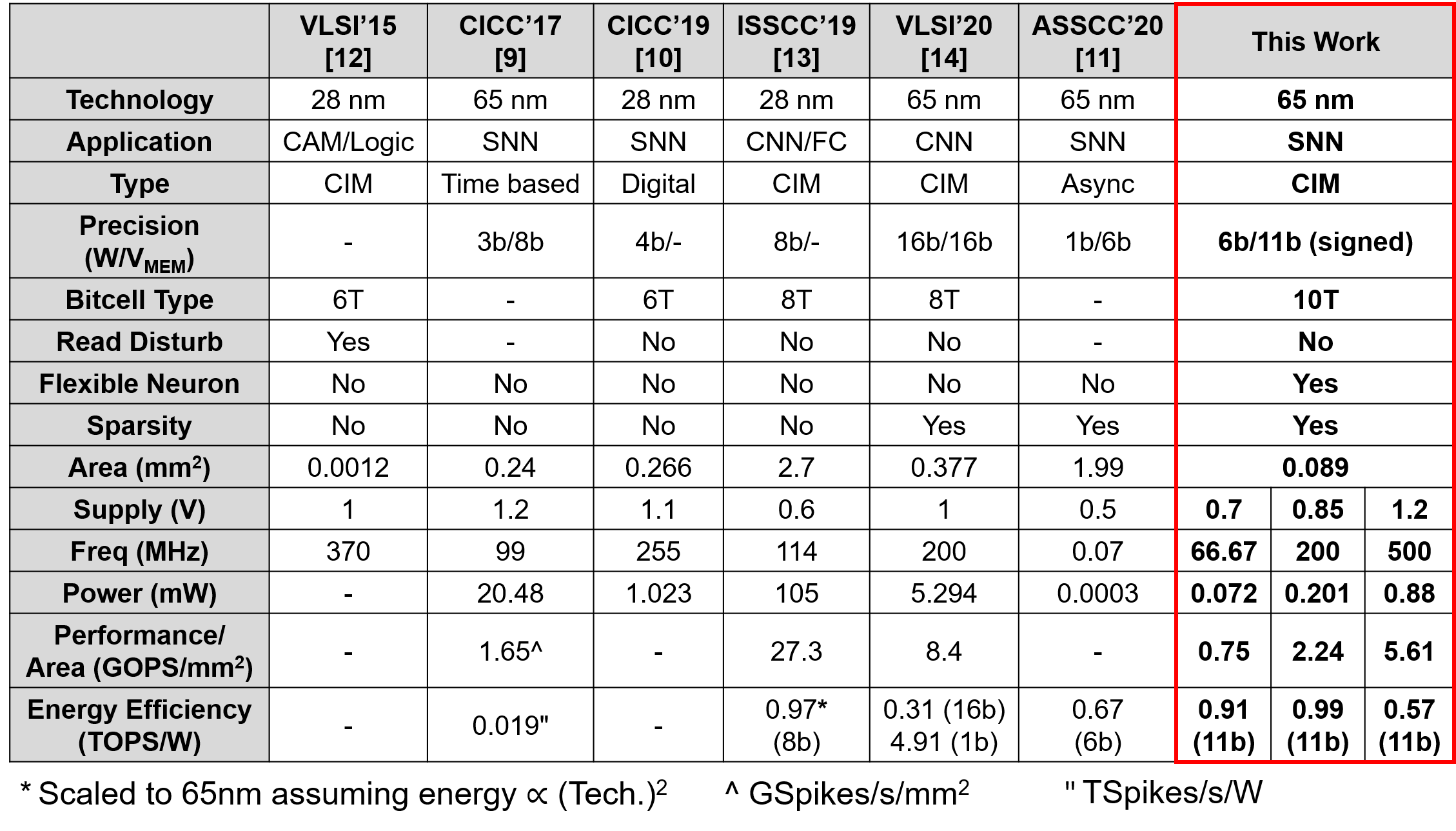}
    \vspace{-9mm}
    \label{fig:comp_table}
\end{table}

\vspace{-1mm}
\section{Conclusion}

In this paper, we present a digital CIM macro with fused weights and membrane potential, designed for efficient processing of SNN inference. The proposed macro inherently leverages the sparsity and supports multiple neuron functionalities. An optimal energy-efficiency of 0.99 TOPS/W was achieved for 11-bit signed operations. We demonstrate our approach by training an SNN for a sentiment analysis task utilizing the intrinsic dynamics of V\textsubscript{MEM}, and also for an image classification task, achieving competitive accuracy to their corresponding LSTM and ANN counterpart, respectively.

\vspace{-1.5mm}
\section*{Acknowledgements}
\vspace{-1mm}
The research was funded in part by C-BRIC, one of six centers in JUMP, a Semiconductor Research Corporation (SRC) program sponsored by DARPA.
\vspace{-1.5mm}

\bibliographystyle{IEEEtran}
\bibliography{ref}

\begin{thebibliography}{10}
\providecommand{\url}[1]{#1}
\csname url@samestyle\endcsname
\providecommand{\newblock}{\relax}
\providecommand{\bibinfo}[2]{#2}
\providecommand{\BIBentrySTDinterwordspacing}{\spaceskip=0pt\relax}
\providecommand{\BIBentryALTinterwordstretchfactor}{4}
\providecommand{\BIBentryALTinterwordspacing}{\spaceskip=\fontdimen2\font plus
\BIBentryALTinterwordstretchfactor\fontdimen3\font minus
  \fontdimen4\font\relax}
\providecommand{\BIBforeignlanguage}[2]{{%
\expandafter\ifx\csname l@#1\endcsname\relax
\typeout{** WARNING: IEEEtran.bst: No hyphenation pattern has been}%
\typeout{** loaded for the language `#1'. Using the pattern for}%
\typeout{** the default language instead.}%
\else
\language=\csname l@#1\endcsname
\fi
#2}}
\providecommand{\BIBdecl}{\relax}
\BIBdecl

\bibitem{davies2018loihi}
M.~Davies \emph{et~al.}, ``Loihi: A neuromorphic manycore processor with
  on-chip learning,'' \emph{IEEE Micro}, vol.~38, no.~1, pp. 82--99, 2018.

\bibitem{merolla2014million}
P.~A. Merolla \emph{et~al.}, ``A million spiking-neuron integrated circuit with
  a scalable communication network and interface,'' \emph{Science}, vol. 345,
  no. 6197, pp. 668--673, 2014.

\bibitem{rathi2020diet}
N.~Rathi \emph{et~al.}, ``Diet-snn: Direct input encoding with leakage and
  threshold optimization in deep spiking neural networks,'' \emph{arXiv
  preprint arXiv:2008.03658}, 2020.

\bibitem{hochreiter1997long}
S.~Hochreiter \emph{et~al.}, ``Long short-term memory,'' \emph{Neural
  computation}, vol.~9, no.~8, pp. 1735--1780, 1997.

\bibitem{narayanan2020spinalflow}
S.~Narayanan \emph{et~al.}, ``Spinalflow: an architecture and dataflow tailored
  for spiking neural networks,'' in \emph{2020 ACM/IEEE 47th Annual
  International Symposium on Computer Architecture (ISCA)}, pp. 349--362.

\bibitem{han2020rmp}
B.~Han \emph{et~al.}, ``Rmp-snn: Residual membrane potential neuron for
  enabling deeper high-accuracy and low-latency spiking neural network,'' in
  \emph{Proceedings of the 2020 IEEE/CVF Conference on Computer Vision and
  Pattern Recognition, {CVPR}}, pp. 13\,558--13\,567.

\bibitem{maas2011learning}
A.~Maas \emph{et~al.}, ``Learning word vectors for sentiment analysis,'' in
  \emph{Proceedings of the 2011 annual meeting of the association for
  computational linguistics: Human language technologies}, pp. 142--150.

\bibitem{pennington2014glove}
J.~Pennington \emph{et~al.}, ``Glove: Global vectors for word representation,''
  in \emph{Proceedings of the 2014 conference on empirical methods in natural
  language processing (EMNLP)}, pp. 1532--1543.

\bibitem{liu2017scalable}
M.~Liu \emph{et~al.}, ``A scalable time-based integrate-and-fire neuromorphic
  core with brain-inspired leak and local lateral inhibition capabilities,'' in
  \emph{2017 IEEE Custom Integrated Circuits Conference (CICC)}, pp. 1--4.

\bibitem{koo2019area}
J.~Koo \emph{et~al.}, ``Area-efficient transposable 6t sram for fast online
  learning in neuromorphic processors,'' in \emph{2019 IEEE Custom Integrated
  Circuits Conference (CICC)}, pp. 1--4.

\bibitem{9336139}
D.~{Wang} \emph{et~al.}, ``Always-on, sub-300-nw, event-driven spiking neural
  network based on spike-driven clock-generation and clock- and power-gating
  for an ultra-low-power intelligent device,'' in \emph{2020 IEEE Asian
  Solid-State Circuits Conference (A-SSCC)}, pp. 1--4.

\bibitem{jeloka2015configurable}
S.~Jeloka \emph{et~al.}, ``A configurable tcam/bcam/sram using 28nm push-rule
  6t bit cell,'' in \emph{2015 IEEE Symposium on VLSI Circuits}, pp.
  C272--C273.

\bibitem{wang201914}
J.~Wang \emph{et~al.}, ``14.2 a compute sram with bit-serial
  integer/floating-point operations for programmable in-memory vector
  acceleration,'' in \emph{2019 IEEE International Solid-State Circuits
  Conference-(ISSCC)}, pp. 224--226.

\bibitem{kim2020z}
J.-H. Kim \emph{et~al.}, ``Z-pim: An energy-efficient sparsity aware
  processing-in-memory architecture with fully-variable weight precision,'' in
  \emph{2020 IEEE Symposium on VLSI Circuits}, pp. 1--2.

\end{thebibliography}

\end{document}